\documentclass[a4paper,12pt]{article}
\RequirePackage{ifpdf}
\usepackage{soul}
\usepackage[usenames,dvipsnames]{color}
\usepackage{jheppub}
\usepackage{esvect}
\usepackage{youngtab}
\usepackage{bbold}
\usepackage{float}
\usepackage{hyperref}

\begin{document}

\title{Defect Partition Function from TDLs in Commutant Pairs}
	\author{Subramanya Hegde\footnote{Corresponding author},}
	\author{Dileep P. Jatkar}
	
	\affiliation{Harish-Chandra Research Institute, Homi Bhabha National Institute}
	\affiliation{Chhatnag Road, Jhunsi, Allahabad, India - 211019}
	\emailAdd{subramanyahegde@hri.res.in}
	\emailAdd{dileep@hri.res.in}

	\abstract{We study topological defect lines in two character
          rational conformal field theories.  Among them one set of
          two character theories are commutant pairs in $E_{8,1}$
          conformal field theory.  Using these defect lines we
          construct defect partition function in the $E_8$ theory.  We
          find that the defects preserve only a part of the $E_8$
          current algebra symmetry.  We also determine the defect
          partition function in $c=24$ CFTs using these defects lines
          of 2 character theories and we find that, with appropriate
          choice of commutant pairs, these defects preserve all
          current algebra symmetries of c = 24 CFTs.}
	
	\maketitle
	

\section{Introduction and summary}
\label{sec:intro}
Two-dimensional conformal field theories (2DCFT) have played a pivotal
role in understanding a variety of problems in theoretical physics,
ranging from string theory \cite{DiFrancesco:1997nk} to mesoscopic
physics \cite{Oshikawa:1996dj} to quantum information \cite{pachos2012}.
All these applications, in turn, have helped deepen our understanding of
2DCFT.  However, classification and study of 2DCFTs is an interesting
problem in its own right \cite{DiFrancesco:1997nk}.  A programme of
classifying 2D rational conformal field theories (RCFTs) by the number
of characters was addressed long ago 
\cite{Anderson:1987ge,Mathur:1988na}.  Mathur, Mukhi, and Sen(MMS)
used the technique of modular differential equations to classify
theories with 2 and 3 characters \cite{Mathur:1988na}.

With renewed interest in the bootstrap programme, Mukhi and
collaborators revisited the classification program of the RCFTs
\cite{Hampapura:2015cea,Gaberdiel:2016zke,Mukhi:2019xjy,Mukhi:2020sxt}.
The MMS classification uses modular linear differential equation
method for classifying admissible charactars.  The solutions depend on
an integer $\ell$, which counts the zeros of the Wronskian of the
solutions.  For two character theories, MMS classification gives
complete solution for $\ell=0$.  The set for $\ell=2$ forms a
commutant pair, in $c=24$ meromorphic CFTs, of the MMS classification.
For higher values of $\ell$ classification of admissible characters is
more involved.  Following the work of Harvey and
Wu\cite{Harvey:2018rdc} which gave a prescription of constructing
higher $\ell$ characters using Hecke images of $\ell=0$ characters,
Chandra and Mukhi\cite{Chandra:2018pjq} showed that admissible
characters for $\ell\ge 6$ can be constructed using appropriate linear
combination of $\ell=0$ quasi-characters\footnote{Quasi-characters are
      q-series with integer coefficients which are not necessarily
      positive.}. Whether these admissible characters can be
associated to RCFTs remains an open problem\cite{Mukhi:2019xjy}.
Recently the program of classification of 2 character RCFTs has been
extended up to the central charge $c<25$\cite{Mukhi:2022bte}.

In this note, we will be focusing on 2-character RCFTs and in fact
will consider a subset of them which belong to MMS series and its
commutants in a single character theory with $c=24$.  We will study
topological defect line(TDL) \cite{Chang:2018iay} in these theories and
construct defect partition functions.  Within the MMS series of
2-character theories, we study TDLs in each of these conformal field
theories.  In general, the TDLs can be invertible if they are
associated with some global
symmetry \cite{Chang:2018iay,Lin:2019kpn,Lin:2019hks}.  On the other
hand, non-symmetry defects are typically non-invertible.  Even within
the MMS series we encounter TDLs of both type.  While in case of the
commutant pairs like ($A_1$, $E_7$), ($A_2$, $E_6$), or ($D_4$,
$D_4$), we get invertible defects, in case of ($G_2$, $F_4$) and
(Lee-Yang(L-Y), $E_{7.5}$) we have non-invertible defects.

It is known that the $E_{8,1}$ character can be decomposed in terms of
sums of products of characters of the commutants.  For example,
\begin{equation}
\label{eq:1}
\chi^{E_{8,1}} = \chi_0^{A_{1,1}}\tilde{\chi}_0^{E_{7,1}} + \chi_{1/4}^{A_{1,1}}\tilde{\chi}_{3/4}^{E_{7,1}}\ .
\end{equation}
Using the TDLs in the commutant pairs, we construct the defect
character in $E_{8,1}$ CFT.  We hasten to emphasise that this is a
non-trivial result since $E_{8,1}$ is a single character theory and
therefore it has no Verlinde lines of its own.  All the TDLs in the
commutant pairs, whether invertible or not, do not commute with the
symmetry currents of $E_8$.  This in turn implies that the level 1
degeneracy of the defect character is different from that of the
$E_{8,1}$ character.  The defect partition function of the $E_{8,1}$ theory is,
\begin{align}\label{defect-partfn-17}
(Z_\eta)^{E_8}=|\chi_0\tilde{\chi}_\frac{3}{4}+\chi_\frac{1}{4}\tilde{\chi}_0|^2\ .
\end{align}
This is one of the main results of this paper.  We compute the defect
partition function using other commutant pairs within the MMS series
as well.  We also give group theoretic interpretation of the defect
partition function.

We also discuss commutant pairs of the MMS series of RCFTs in $c=24$
CFTs except L-Y and $E_{7.5}$.  All the relevant $c=24$ CFTs possess
current algebra symmetry.  We find that whenever we have a symmetry
defect, it preserves the entire current algebra symmetry of the $c=24$
CFT.  This is in sharp contrast with what happens in the $E_{8,1}$
case where the defect preserves only a part of the $E_8$ symmetry.
The origin of this difference lies in the fact that the dimensions of
non-trivial primary operators in a given commutant pair of $E_8$ add up
to 1. Hence the defect in the commutant component affects the level 1
degeneracy of the $E_8$ theory. On the other hand, the commutant pairs in
c = 24 CFTs, that we will be focussing on in this manuscript, has the
dimensions of non-identity primary adding up to
2\cite{Gaberdiel:2016zke}. As a result the defect in the commutant
component leaves the level 1 degeneracy invariant.

This note is organised as follows: In section 2, we briefly recall the
aspects of topological defects.  We focus on the TDLs associated with
global symmetries as well as those without any symmetry.  In section
3, we take up the MMS series and study the TDLs in each CFT in detail.
We also show that these defects can be embedded in the $E_8$ theory,
and the defect character in the $E_8$ theory is computed using the
modular S transformation of the character with the defect insertion.
In section 4, we briefly discuss defects in MMS CFTs and their
commutants in $c=24$ CFTs.  These theories have been discussed in 
\cite{Mukhi:2019xjy, Gaberdiel:2016zke,Hampapura:2015cea}, and their
$c=24$ parent belongs to the Schellekens
classification \cite{Schellekens:1992db}.  Although our analysis can be
extended to the entire series we focus only on two cases, namely $A_1$
and its commutant in $c=24$, and $E_7$ and its commutant in $c=24$.
We conclude this note by summarising our results.

\section{Topological defects}

Topological defects are generalisations of global symmetries. Consider
a continuous global symmetry in a $d$ dimensional field theory, the
corresponding charge is given by $Q_a=\int d^{d-1}x\, n_\mu j^\mu_a$,
where $j^\mu_a$ is the current. Action of the symmetry on the states
is then implemented by, $\mathcal{L}_\theta=\exp(i\theta^aQ_a)$, where
$\theta^a$ are the transformation parameters. In a CFT, the charge and
hence the operator $\mathcal{L}_\theta$ commute with the conformal
generators. The action of the symmetry generators on the state can be
understood through its action the corresponding operator via
state-operator correspondence.  When the co-dimension one surface
defect operator $\mathcal{L}_\theta$ encloses an operator, it leads to
a new operator in the CFT.  Since the defect commutes with the
conformal generators, the conformal dimension of the operator is
preserved. In a two-dimensional CFT, the line operator(which is a
co-dimension one defect) commutes with the Virasoro generators and
hence can be deformed as long as there are no other insertions around
them.  Therefore, the operation is termed as acting on by a topological
defect line (TDL). The operators $\mathcal{L}_g$ are associated with
symmetry elements $g$ for continuous as well as discrete global
symmetries. TDLs associated with global symmetries are known as
invertible defect lines, as the operators on the endpoint of defects
have inverses under the fusion rules \cite{Chang:2018iay}. There are
several examples of TDL that are not invertible and hence do not
correspond to symmetries of the theory. For example, duality defect in
the Ising CFT \cite{frohlich2004kramers}.

In diagonal modular invariant RCFTs, there is a TDL associated with
each primary of the theory as \cite{Petkova:2000ip, Verlinde:1988sn},
\begin{align}\label{verlinde-line-projetor}
\mathcal{L}_i=\sum_k\frac{S_{ik}}{S_{i0}}\mathcal{P}_k,
\end{align}
where $S_{ij}$ are the elements of the symmetric and unitary modular
$S$ matrix for the transformation of the characters, and
$\mathcal{P}_k$ is the projector to the module generated by $k$. It is
important to note that the above operators commute with not only the
Virasoro algebra but the full chiral algebra of the theory. TDLs given
as above are known as Verlinde lines \cite{Petkova:2000ip,
  Verlinde:1988sn, Drukker_2011, Gaiotto:2014lma}. Their fusion
follows that of the corresponding primaries according to the Verlinde
formula.

Given a Verlinde line, one can define partition functions with Verlinde lines along different non-contractible cycles. Defect lines running along the space directions are encountered while taking the trace to compute the partition function as insertions of operators. Defect lines running along the time direction impose a boundary condition on each spatial slice and hence modify the Hilbert space over which the trace is performed. Given a Verlinde line of the form \eqref{verlinde-line-projetor}, we can obtain the partition function with a defect insertion as,
\begin{align}\label{insertion-partition-function}
\mathcal{Z}^{\mathcal{L}_i}=Tr(\mathcal{L}_ie^{-q(L_0-\frac{c}{24})})=\sum_k\frac{S_{ik}}{S_{i0}}\bar{\chi}_k(\bar{\tau})\chi_k(\tau).
\end{align}
We can obtain the partition function with the defect line inserted along the time direction by a modular $S$ transformation as  \cite{Petkova:2000ip}, 
\begin{align}
\mathcal{Z}_{\mathcal{L}_i}(\tau,\bar{\tau})&=\mathcal{Z}^{\mathcal{L}_i}(-\frac{1}{\tau},-\frac{1}{\bar{\tau}})\nonumber\\
&=\sum_{k,l,m}\frac{S^*_{kl}S_{ik}S_{km}}{S_{i0}}\bar{\chi}_l\chi_m\nonumber\\
&=\sum_{l,m}N_{im}^l\bar{\chi}_l\chi_m,
\end{align}
where $N_{im}^l=\sum_k\frac{S^*_{kl}S_{ik}S_{km}}{S_{i0}}$ are the
fusion coefficients given by the Verlinde formula and are
integers. They give the degeneracy of the operators in the defect
Hilbert space with dimensions $(\bar{h}_l,h_m)$.  Thus one can read
off the operator content in the defect Hilbert space by expressing
$\mathcal{Z}_\mathcal{L}$ in terms of the characters. This utility
extends beyond the Verlinde lines to more general TDLs.

While partition functions for defects that commute with the full chiral algebra of the theory have been studied extensively in the literature, recently more general defects have been considered \cite{Lin:2019hks}. In  \cite{Lin:2019hks}, the authors considered defects inside the Monster CFT which do not commute with the Monster group.  A key component in the analysis was that the character for the Monster CFT can be expressed in terms of Ising and Baby Monster CFT characters as,
\begin{align}\label{monster-ising-baby}
\chi^{\text{M}}=\chi_0^{\text{Ising}}\chi_0^{\text{Baby}}+\chi_{\frac{1}{2}}^{\text{Ising}}\chi_\frac{3}{2}^{\text{Baby}}+\chi_{\frac{1}{16}}^{\text{Ising}}\chi_\frac{31}{16}^{\text{Baby}},
\end{align}
where $\chi^M=j(\tau)-744$ is the character for the identity module of the Monster CFT, and on the RHS the chiral characters are those of Ising modules for primaries with dimensions $0,\frac{1}{2},\frac{1}{16}$ and Baby Monster primaries with dimensions  $0,\frac{3}{2},\frac{31}{16}$. It was shown that the above relation between the characters can be interpreted in terms of fermionisation of the Monster theory. In particular, it was shown that one can identify the fermionised Monster CFT as a tensor product of the fermionised Baby Monster and the Majorana-Weyl CFT. This allowed the realisation of various defects of the Ising category inside Monster CFT. In particular, the duality defect of the bosonic Monster CFT was implemented as a $\mathbb{Z}_2$ defect of the fermionic Monster where the $\mathbb{Z}_2$ corresponds to a sign flip of the Majorana-Weyl fermion. A useful way indicated in \cite{Lin:2019hks} to obtain the partition functions for Monster CFT with defects running along different non-contractible cycles is the following. From  \eqref{monster-ising-baby}, note that as the LHS is modular invariant, the $S$ matrix for the transformation of the Ising CFT and Baby Monster CFT are identical. Therefore, given the partition function with an insertion of the $\mathbb{Z}_2$ Verlinde line corresponding to the primary with dimension $1/2$,
\begin{align}
\mathcal{Z}^{\text{Ising}\,\eta}=\overline{\chi_0^{\text{Ising}}}\chi_0^{\text{Ising}}+\overline{\chi_\frac{1}{2}^{\text{Ising}}}\chi_\frac{1}{2}^{\text{Ising}}-\overline{\chi_\frac{1}{16}^{\text{Ising}}}\chi_\frac{1}{16}^{\text{Ising}},
\end{align}
we can write the corresponding partition function with insertion in
the Monster CFT as,
\begin{align}
\mathcal{Z}^{M\eta}=\chi_0^{\text{Baby}}\chi_0^{\text{Ising}}+\chi_\frac{3}{2}^{\text{Baby}}\chi_{\frac{1}{2}}^{\text{Ising}}-\chi_\frac{31}{16}^{\text{Baby}}\chi_{\frac{1}{16}}^{\text{Ising}},
\end{align}
where each $\bar{\chi}^{\text{Ising}}$ is replaced by the
corresponding $\chi^{\text{Baby}}$. As the modular $S$ matrices are
identical, the partition function for the Monster has the appropriate
modular transformation property.  In particular, the same replacement
rule holds for the defect running along the time direction. It is
important to note that the defects $\eta$ and $N$ of the Ising
category realised inside the Monster CFT in this way do not commute
with the full Monster group. They preserve only a part of the chiral
algebra of the CFT. Bosonisation of the fermionic Baby Monster CFT
extends the symmetry to its double cover $2.\mathbb{B}$. The defects
commute with this double cover, which is a subgroup of the Monster
group. The elements in the defect partition function could be
decomposed in terms of the dimensions of Baby Monster representations.

In the following section, we will consider partition functions with defect insertions realised in the $E_{8,1}$ CFT and various $c=24$ meromorphic CFTs by an analogous replacement rule. We will see that the defects defined using such a rule commute with a subgroup of the chiral algebra of the theory.

\section{Defects in Mathur-Mukhi-Sen conformal field theories}
In this section we will consider CFTs that appear in the
Mathur-Mukhi-Sen series of two character rational CFTs 
\cite{Mathur:1988na}. These CFTs form a pair inside the $E_{8,1}$ WZW
model by the relation
\begin{align}
\chi^{E_{8,1}}=\chi_0\tilde{\chi}_0+N\chi_1\tilde{\chi}_1,
\end{align}
where $N$ is an integer that denotes the degeneracy of the non-identity
primary. The characters $\chi$ and $\tilde{\chi}$ correspond to pairs
inside the series whose central charges add to $8$ and conformal
dimensions of the non-identity primary add to $1$. The $E_{8,1}$ WZW is
a single character CFT, therefore the modular $S$ matrices of the pair
are either identical, or complex conjugates of each other.  In either
case the structure of the partition functions with Verlinde lines is
identical. We will use this to define defects inside the $c=8$ theory
by using a replacement rule, which will be explained in the next
sub-section.  For the cases of $A_{1,1}/E_{7,1}$ and $A_{2,1}/E_{6,1}$, we
will illustrate, using the branching rules for $E_8$, the action of the
defect and determine the subgroup of $E_8$ that commute with the defect.

The MMS series CFTs have the following expressions for their characters  \cite{Mukhi:2020sxt,Gaberdiel:2016zke} which we will use in this section.
\begin{align}
	\chi_{0}(\tau)&=j^\frac{c}{24}{ }_{2} F_{1}(\frac{1}{12}-\frac{h}{2},\frac{5}{12}-\frac{h}{2}, 1-h ;\frac{1728}{j})\nonumber\\
	\chi_{h}(\tau)&=\frac{|\sqrt{m}|}{\sqrt{N}}j^{\frac{c}{24}-h}_{2} F_{1}(\frac{1}{12}+\frac{h}{2},\frac{5}{12}+\frac{h}{2}, 1+h ;\frac{1728}{j}),
\end{align}
where
\begin{align}	\sqrt{m}=(1728)^h\left(\frac{\sin(\pi(\frac{1}{12}-\frac{h}{2}))
\sin(\pi(\frac{5}{12}-\frac{h}{2}))}{\sin(\pi(\frac{1}{12} +\frac{h}{2}))\sin(\pi(\frac{5}{12}+\frac{h}{2}))}\right)^{0.5} \frac{\Gamma(1-h)\Gamma(\frac{11}{12}+\frac{h}{2}) \Gamma(\frac{7}{12}+\frac{h}{2})}{\Gamma(1+h) \Gamma(\frac{11}{12}-\frac{h}{2})\Gamma(\frac{7}{12}-\frac{h}{2})}.
\end{align} 
The factor $\frac{|\sqrt{m}|}{\sqrt{N}}$ is an integer for each of these theories.

\subsection{$A_{1,1}$ CFT/ $E_{7,1}$ CFT}

The pair of CFTs here are $A_{1.1}$ and $E_{7,1}$ WZW model with
central charge $c=1$ and $c=7$ respectively. As we will explain later
in this section, they have the same modular S matrix given by 
\cite{DiFrancesco:1997nk},
\begin{align}
S=\frac{1}{\sqrt{2}}\begin{pmatrix}
1 & 1\\
1 & -1
\end{pmatrix}.
\end{align}
Partition function for the $A_{1,1}$ theory is given as,
\begin{align}
\mathcal{Z}^{A_{1,1}}=\overline{\chi^{A_{1,1}}_0}\chi^{A_{1,1}}_0+\overline{\chi^{A_{1,1}}_{\frac{1}{4}}}\chi^{A_{1,1}}_{\frac{1}{4}},
\end{align}
where $\chi^{A_{1,1}}_0,\chi^{A_{1,1}}_{\frac{1}{4}}$ (or $\overline{\chi^{A_{1,1}}_0},\overline{\chi^{A_{1,1}}_{\frac{1}{4}}}$)  are the (anti-)chiral characters corresponding to
Ka\v{c}-Moody primaries with scaling dimensions $h_L=h_R=0$ and
$h_L=h_R=\frac{1}{4}$ respectively. The non-identity primary transforms under the $\mathbf{2}$ representation of $A_{1,1}$. The $q$ expansion of the characters read,
\begin{align}\label{q-expansion-A11}
\chi^{A_{1,1}}_0&=q^{-\frac{1}{24}}(1+3q+4q^2+7q^3+\cdots)\ ,\nonumber\\
\chi^{A_{1,1}}_{\frac{1}{4}}&=q^{\frac{5}{24}}(2+2q+6q^2+8q^3+\cdots)\, .
\end{align}
Let us consider the Verlinde line corresponding to the non-identity
primary. The action of the Verlinde line is given by,
\begin{align} 
\mathcal{L}_2|\phi_0\rangle
&=\frac{S_{10}}{S_{00}}|\phi_0\rangle= |\phi_0\rangle,\nonumber\\
\mathcal{L}_2|\phi_{\frac{1}{4}}\rangle
&=\frac{S_{11}}{S_{01}}|\phi_{\frac{1}{4}}\rangle=-|\phi_{\frac{1}{4}}\rangle.
\end{align}
Therefore the partition function with $\mathcal{L}_2$ insertion is given according to \eqref{insertion-partition-function} by,
\begin{align}\label{etainsertionsu2}
\mathcal{Z}^{A_{1,1}\,\mathcal{L}_2}=|\chi_0|^2-|\chi_{\frac{1}{4}}|^2.
\end{align}
Note that as the insertion keeps the vacuum invariant and flips the sign for the fundamental representation, the $\mathbb{Z}_2$ can be interpreted as the center of $A_1$. This has been discussed recently in  \cite{Lin:2021udi}. We will come back to this later in this section.  

To obtain the defect Hilbert space partition function, where the defect runs along the time direction, we perform a
$S$ modular transformation as explained in the previous section. We obtain,
\begin{align}\label{etaparitionfunctionsu2}
\mathcal{Z}^{A_{1,1}}_{\mathcal{L}_2}(\tau)
&=\mathcal{Z}^{A_{1,1}\,\mathcal{L}_2}(-1/\tau)\nonumber\\
&=|\chi_0(-1/\tau)|^2-|\chi_{\frac{1}{4}}(-1/\tau)|^2\nonumber\\
&=\bar{\chi}_0(\tau)\chi_{\frac{1}{4}}(\tau)+\bar{\chi}_{\frac{1}{4}}(\tau)\chi_0(\tau).
\end{align}
Therefore, the scaling dimensions of the operators in the defect
Hilbert space is given as $(h_L,h_R)=(0,\frac{1}{4})$ and
$(h_L,h_R)=(\frac{1}{4},0)$. Using further modular $T$ transformations of the defect partition function, one can diagnose the 't Hooft anomaly for this symmetry defect. This was studied recently in  \cite{Chang:2018iay,Lin:2019kpn,Lin:2021udi,hung2014universal}. The partition function $\mathcal{Z}^{A_{1,1}\,\mathcal{L}_2}(\tau)$ is invariant under the congruent subgroup $\Gamma_0(4)$ of $SL(2,\mathbb{Z})$. Modular properties of the defects in the MMS series can be inferred from the results in \cite{Mathur:1989pk}.

For $E_7$ WZW model, the primaries are of dimension $h_L=h_R=0$ and
$h_L=h_R=\frac{3}{4}$. The non-identity primary trasnforms under the $\mathbf{56}$ representation of $E_{7,1}$. We will denote the corresponding holomorphic characters as $\tilde{\chi}^{E_{7,1}}_0$ and $\tilde{\chi}^{E_{7,1}}_{\frac{3}{4}}$(and similarly for antiholomorphic characters), whose $q$ expansions are given as,
\begin{align}\label{q-expansion-E71}
\tilde{\chi}^{E_{7,1}}_0&=q^{-\frac{7}{24}}(1+133q+1673q^2+11914q^3+\cdots)\nonumber\\
\tilde{\chi}^{E_{7,1}}_{\frac{3}{4}}	&=q^{\frac{11}{24}}(56+968q+7504q^2+42616q^3+\cdots).
\end{align}

As discussed in the beginning of this section, $A_{1,1}$ and $E_{7,1}$ WZW satisfy,
\begin{align}
\chi_0^{E_{8,1}}=\chi^{A_{1,1}}_0\tilde{\chi}^{E_{7,1}}_0+\chi^{A_{1,1}}_{\frac{1}{4}}\tilde{\chi}^{E_{7,1}}_{\frac{3}{4}},
\end{align}
where $\chi_0^{E_{8,1}}$ is the character for $E_{8,1}$ WZW model whose $q$ expansion reads,
\begin{align}
\chi_0^{E_{8,1}}=q^{-\frac{1}{3}}(1+248q+4124q^2+34752q^3+\cdots).
\end{align}
The above expression can be made sense of by using the branching rules for $E_8$  representations into $A_1$ and $E_7$ representations. Let us note the following branching rules \cite{Slansky:1981yr}. 
\begin{align}
\mathbf{(248)}&=\mathbf{(3,1)}+\mathbf{(1,133)}+\mathbf{(2,56)},\nonumber\\
\mathbf{(3875)}&=\mathbf{(1,1)}+\mathbf{(2,56)}+\mathbf{(3,133)}+\mathbf{(1,1539)}+\mathbf{(2,912)}.
\end{align}
We can then write the coefficients as,
\begin{align}
248&=3\times 1+(1\times 133)+(2\times 56)\nonumber\\
4124&=1+248+(1\times 1)+(2\times 56)+(3\times 133)+(1\times 1539)+(2\times 912).
\end{align}
Note that as $E_{8,1}$ WZW model is a single character theory, its character is invariant under the modular $S$ transformation. Therefore, the modular $S$ matrix for $E_{7,1}$ theory should be the inverse transpose of the modular $S$ matrix of $A_{1,1}$ theory.

However, modular $S$-matrix of $A_{1,1}$ is a symmetric orthogonal matrix. Therefore, modular $S$ matrix of the two theories are identical. Thus the partition functions with defect $\mathcal{L}_{56}$ for $E_{7,1}$ have the same structure as the $\mathcal{L}_{2}$ partition functions considered above if we replace $\chi_0$ with $\tilde{\chi}_0$ and $\chi_{\frac{1}{4}}$ with $\tilde{\chi}_\frac{3}{4}$. Therefore the scaling dimensions of the operators in the defect
Hilbert space is given as $(h_L,h_R)=(0,\frac{3}{4})$ and
$(h_L,h_R)=(\frac{3}{4},0)$. Again, this defect can be interpreted as the action by the center of $E_{7,1}$. 

The partition function of $E_{8,1}$ WZW model in terms of the characters defined above is,
\begin{align}
  \label{eq:2}
Z^{E_{8,1}}&=|\chi_0^{E_{8,1}}|^2\nonumber\\
&=|\chi^{A_{1,1}}_0\tilde{\chi}^{E_{7,1}}_0+\chi^{A_{1,1}}_{\frac{1}{4}}\tilde{\chi}^{E_{7,1}}_{\frac{3}{4}}|^2.
\end{align}
Using \eqref{eq:2} and \eqref{etainsertionsu2}, we are motivated to define the following partition function for $E_{8,1}$ WZW model with an $\eta$ insertion,
\begin{align}\label{insertion-partition-function-17}
(Z^{\eta_{1,7}})^{E_{8,1}}=|(\chi^{\eta_{1,7}})^{E_{8,1}}|^2\equiv|\chi_0^{A_{1,1}}\tilde{\chi}^{E_{7,1}}_0-\chi^{A_{1,1}}_\frac{1}{4}\tilde{\chi}^{E_{7,1}}_\frac{3}{4}|^2,
\end{align}
which we obtained by using the RHS of \eqref{etainsertionsu2} and
replacing the $\bar{\chi}$ in each term with the corresponding
$\tilde{\chi}$ and then taking the modulus squared of the
expression. In the rest of the paper, this procedure will be referred to as the `replacement rule'.  We have chosen to denote the defect as $\eta_{1,7}$ where
the subscript stands for the central charges of the $A_{1,1}$ and
$E_{7,1}$ pair.

The above twisted character $(\chi^{\eta_{1,7}})^{E_{8,1}}$ has the following $q$ series expansion,
\begin{align}
(\chi^{\eta_{1,7}})^{E_{8,1}}&=q^{-\frac{1}{3}}(1+24q+28q^2+192q^3+\cdots)
\end{align}
Recall that we interpreted the defect as the one generated by the
center of $A_{1,1}$. We already know that the fundamental
representation changes sign under the action of the center and hence
the defect. We use the Clebsch-Gordan formula to identify which
representations change sign. For $A_{1,1}$, as expected, all the even
dimensional representations change sign. We then have the decomposition of above
coefficients as,
\begin{align}
24&=3\times 1+(1\times 133)-(2\times 56)\nonumber\\
28&=1+24+(1\times 1)-(2\times 56)+(3\times 133)+(1\times 1539)-(2\times 912).
\end{align}
Thus we can interpret the operator $\eta_{1,7}$ inside $E_{8,1}$ as
the center of $A_{1,1}$. Note that we can also interpret the action of
$\eta_{1,7}$ as the center of $E_{7,1}$ instead but not both. This is
so because $E_8$ has a maximal subgroup $(A_1 \otimes
E_7)/(-1,-1)$. Therefore the non trivial element in the center of
$A_1$ and $E_7$ are identified due to the $(-1,-1)$
identification. Thus the defect operator $\eta_{1,7}$ inside $E_{8,1}$
is the equivalence class of $\{\{1,-1\},\{-1,1\}\}$ acting on
appropriate representations inside $A_1$ and $E_7$ as per the above
decomposition. This element, of course, does not commute with the full
$E_8$. When one expresses $E_8$ as
$(A_1 \otimes E_7) / (-1,-1)+ 2 \otimes 56$, then this decomposition
has a non-algebraic double cover. The defect commutes with this non
algebraic double cover inside $E_8$.

Using the modular $S$ transformation of the characters $\chi$ and
$\tilde{\chi}$, we find the defect partition function,
\begin{align}\label{defect-hilbert-part-17}
(Z_\eta)^{E_8}=|\chi_0\tilde{\chi}_\frac{3}{4}+\chi_\frac{1}{4}\tilde{\chi}_0|^2
\end{align}
This is consistent, if we take \eqref{etaparitionfunctionsu2} and
follow the same replacement rule as we have used above to obtain the
partition function with the insertion. From the above formula, we can
deduce the modular transformation property of the defect insertion
partition function \eqref{insertion-partition-function-17}. We see
that \eqref{defect-hilbert-part-17} given above is invariant under
$T^2$ modular transformation. Therefore,
\eqref{insertion-partition-function-17} is invariant under $ST^2S$
which belongs to the congruent subgroup $\Gamma_0(2)$. It can be
checked that the invariance extends to the full $\Gamma_0(2)$.

\subsection{$A_{2,1}$ CFT/$E_{6,1}$ CFT}
The commutant pair $A_{2,1}$ and $E_{6,1}$ WZW model CFTs have central charge $c=2$ and $c=6$ respectively. 

Let us begin with the $A_{2,1}$ theory, the modular S matrix for this
theory is given by \cite{DiFrancesco:1997nk},
\begin{align}
S=\frac{1}{\sqrt{3}}\begin{pmatrix}
1 & 1 & 1\\
1 & -\frac{1}{2}+\frac{i\sqrt{3}}{2} &-\frac{1}{2}-\frac{i\sqrt{3}}{2}\\
1 & -\frac{1}{2}-\frac{i\sqrt{3}}{2} & -\frac{1}{2}+\frac{i\sqrt{3}}{2}
\end{pmatrix}.
\end{align}
The partition function is given as,
\begin{align}
\mathcal{Z}^{A_{2,1}}=\overline{\chi^{A_{2,1}}_0}\chi^{A_{2,1}}_0+2\overline{\chi^{A_{2,1}}_{\frac{1}{3}}}\chi^{A_{2,1}}_{\frac{1}{3}}\ ,
\end{align}
where, $\chi^{A_{2,1}}_0,\chi^{A_{2,1}}_{\frac{1}{3}}$( or
$\overline{\chi^{A_{2,1}}_0},\overline{\chi^{A_{2,1}}_\frac{1}{3}}$)
are (anti-)chiral characters corresponding to Ka\v{c}-Moody primaries
with scaling dimensions $h_L=h_R=0$ and $h_L=h_R=\frac{1}{3}$
respectively. Primaries corresponding to $\mathbf{3}$ and
$\mathbf{\bar{3}}$ representations of $A_{2,1}$ have the same
character $\chi^{A_{2,1}}_{\frac{1}{3}}$. The $q$ expansion of both
$h=0$ and $h=1/3$ characters is given as,
\begin{align}
\chi^{A_{2,1}}_0&=q^{-\frac{1}{12}}(1+8q+17q^2+46q^3+\cdots)\ ,\nonumber\\
\chi^{A_{2,1}}_{\frac{1}{3}}&=3q^{\frac{1}{4}}(1+3q+9q^2+19q^3+\cdots)\ .
\end{align}
Let us consider the Verlinde line corresponding to the non-identity
primary that transforms in the representation $\mathbf{3}$ of
$A_{2,1}$. The action of the Verlinde line on the primary states is,
\begin{align}\label{eq:3}
\mathcal{L}_3|\phi_0\rangle
&=\frac{S_{10}}{S_{00}}|\phi_0\rangle=|\phi_0\rangle,\nonumber\\
\mathcal{L}_3|\phi_{3}\rangle
&=\frac{S_{11}}{S_{01}}|\phi_{3}\rangle=\frac{-1+i\sqrt{3}}{2}|
\phi_{3}\rangle,\nonumber\\
\mathcal{L}_3|\phi_{\bar{3}}\rangle
&=\frac{S_{12}}{S_{02}}|\phi_{\bar{3}}\rangle=\frac{-1-i\sqrt{3}}{2}|
\phi_{\bar{3}}\rangle\ .
\end{align}
The partition function with $\mathcal{L}_3$ insertion takes the form,
\begin{align}\label{insertion-partition-function-A2}
\mathcal{Z}^{A_{2,1}\,\mathcal{L}_3}=|\chi^{A_{2,1}}_0|^2-|\chi^{A_{2,1}}_{\frac{1}{3}}|^2.
\end{align}
It is evident from \eqref{eq:3} that $\mathcal{L}_3$ eigenvalues
belong to cube root of unity and hence its insertion in the partition
function can be interpreted as the action of the center of $A_{2,1}$.

To obtain the defect Hilbert space partition function, we
perform the modular $S$ transformation.
\begin{align}
\mathcal{Z}^{A_{2,1}}_{\mathcal{L}_3}(\tau)
&=\mathcal{Z}^{A_{2,1}\,\mathcal{L}_3}(-1/\tau)\nonumber\\
&=|\chi^{A_{2,1}}_0(-1/\tau)|^2-|\chi^{A_{2,1}}_{\frac{1}{3}}(-1/\tau)|^2\nonumber\\
&=\overline{\chi^{A_{2,1}}_\frac{1}{3}}\chi^{A_{2,1}}_{\frac{1}{3}}+\overline{\chi^{A_{2,1}}_0}
\chi^{A_{2,1}}_{\frac{1}{3}} +\overline{\chi_{\frac{1}{3}}^{A_{2,1}}}\chi^{A_{2,1}}_0.
\end{align}
The scaling dimensions $(h_L,h_R)$ of the operators in the defect
Hilbert space can be read off to be $({\frac{1}{3}},{\frac{1}{3}})$,
$(0,\frac{1}{3})$, and $(\frac{1}{3},0)$ each of them belong to the
representation $\mathbf{3}$ of $A_{2,1}$ but are obtained from
$\mathbf{\bar{3}}\otimes \mathbf{\bar{3}}$,
$\mathbf{1} \otimes \mathbf{3}$ and $\mathbf{3} \otimes \mathbf{1}$
representations of $A_{2,1}$ respectively. Note that there is also an
$\mathcal{L}_{\bar{3}}$ Verlinde line with the same defect Hilbert space partition function.

Let us now consider the $E_{6,1}$ WZW model CFT, the characters of this theory have the $q$ expansion,
\begin{align}
\tilde{\chi}^{E_{6,1}}_0&=q^{-\frac{1}{4}}(1+78q+729q^2+4382q^3+\cdots)\ , \nonumber\\
\tilde{\chi}^{E_{6,1}}_{\frac{2}{3}}&=q^{\frac{5}{12}}(27+378q+2484q^2+12312q^3+\cdots)\ ,
\end{align}
where $\tilde{\chi}^{E_{6,1}}_0,\tilde{\chi}^{E_{6,1}}_{\frac{2}{3}}$
are the characters for the modules corresponding to identity and the
non-identity primaries transforming as $\mathbf{27}$ and
$\mathbf{\overline{27}}$ under $E_{6,1}$.  The modular $S$ matrix for
this model is the complex conjugate of the modular $S$ matrix of the
$A_{2,1}$ model given above.  However it can easily be verified that
the $\mathcal{L}_{27}$ defect partition functions have the same
structure as the $\mathcal{L}_3$ defect partition functions, and are
reproduced on replacing the characters $\chi^{A_{2,1}}$ with the
corresponding characters $\tilde{\chi}^{E_{6,1}}$.

The $E_{8,1}$ CFT character in terms of the commutant pair
$(A_{2,1}, E_{6,1})$ is given as \cite{Mathur:1988na}\footnote{Note
  that we choose to write the degeneracy factor $2$ outside the
  characters. Sometimes the factor is absorbed into the characters as
  $\sqrt{2}$ \cite{Gaberdiel:2016zke,Mukhi:2020sxt}. However for the
  interpretation of the defect Hilbert space partition function this
  will be inconvenient. },
\begin{align}
\chi_0^{E_{8,1}}&=\chi^{A_{2,1}}_0\tilde{\chi}^{E_{6,1}}_0+2\chi^{A_{2,1}}_{\frac{1}{3}}\tilde{\chi}^{E_{6,1}}_{\frac{2}{3}}.
\end{align}
To interpret the decomposition above, we need the branching rules of $E_8$ representations into $A_2\otimes E_6$ representations.  Let us note the following branching rules\cite{Slansky:1981yr},
\begin{align}
\mathbf{(248)}&=\mathbf{(8,1)}+\mathbf{(1,78)}+\mathbf{(3,27)}+\mathbf{(\bar{3},\overline{27})},\nonumber\\
  \mathbf{(3875)}&=\mathbf{(1,1)}+\mathbf{(8,1)}+\mathbf{(\bar{6},27)}+\mathbf{(6,\overline{27})}+\mathbf{(8,78)}+\mathbf{(1,650)}\nonumber\\
  &\hspace{5mm}+\mathbf{(3,351)}+\mathbf{(\bar{3},\overline{351})}.
\end{align}
With the help of these branching rules, it is easy to understand how
the coefficients in the $q$ expansion of the $E_{8,1}$ CFT character $\chi_0^{E_{8,1}}$ decompose in terms of the coefficients in the $q$ expansions of $A_{2,1}$ and $E_{6,1}$ characters, quite in an analogous manner as in the previous subsection.

Using the replacement rule on \eqref{insertion-partition-function-A2}, the twisted chracter for $E_{8,1}$ CFT corresponding to the above defect is given by,
\begin{align}
(\chi^{\eta_{2,6}})^{E_{8,1}}&=\chi^{A_{2,1}}_0\tilde{\chi}^{E_{6,1}}_0-\chi^{A_{2,1}}_\frac{1}{3}\tilde{\chi}^{E_{6,1}}_\frac{2}{3}\nonumber\\
&=q^{-\frac{1}{3}}(1+5q-7q^2+3q^3+\cdots) 
\end{align}
We interpret this as the action of the center of $A_2$ inside $E_8$.  Let us first note that the center of $A_2$ acts with $\omega$ on $\mathbf{3}$ and $\omega^2$ on $\mathbf{\bar{3}}$, where $\omega=\frac{-1+i\sqrt{3}}{2}$ is a cube root of 1. Using the Clebsch-Gordan formula, we can find that $\mathbf{8}$ has trivial phase while $\mathbf{6}$ and $\mathbf{\bar{6}}$ acquire the phase $\omega^2$ and $\omega$ respectively. Thus the coefficients above can be written as,
\begin{align}
5&=(8\times 1)+(1\times 78)+\omega (3\times 37)+\omega^2 (3\times 37),\nonumber\\
&=(8\times 1)+(1\times 78)- (3\times 37),
\end{align} 
and,
\begin{align}
  -7&=1+5+(1\times 1)+(8\times 1)+(\omega+\omega^2)(6\times 27)+(8\times 78)\nonumber\\
  &\hspace{5mm}+(1\times 650)+(\omega+\omega^2)(3\times 351)\nonumber\\
    &=1+5+(1\times 1)+(8\times 1)-(6\times 27)+(8\times 78)\nonumber\\
  &\hspace{5mm}+(1\times 650)-(3\times 351)\ .	
\end{align}
We can also interpret this as the center of $E_{6,1}$ instead, as we did in the previous subsection.

The character with the defect along the time direction is given as,
\begin{align}
(\chi_{\eta_{2,6}})^{E_{8,1}}&=\tilde{\chi}^{E_{6,1}}_\frac{2}{3}\chi^{A_{2,1}}_{\frac{1}{3}}+\tilde{\chi}^{E_{6,1}}_0
\chi^{A_{2,1}}_{\frac{1}{3}} +\tilde{\chi}_{\frac{2}{3}}^{E_{6,1}}\chi^{A_{2,1}}_0.
\end{align} 
The corresponding $(\mathcal{Z}_{\eta_{2,6}})^{E_{8,1}}=|(\chi_{\eta_{2,6}})^{E_{8,1}}|^2$ can be seen to be invariant under $\Gamma_0(3)$. 

\subsection{$G_{2,1}$ CFT/$F_{4,1}$ CFT}
The $G_{2,1}$ and $F_{4,1}$ WZW CFTs have central charge
$c=\frac{14}{5}$ and $c=\frac{26}{5}$ respectively.  Since they form a
commutant pair inside $E_{8,1}$, they share the same modular S matrix,
which is given by \cite{DiFrancesco:1997nk},
\begin{align}
S=\sqrt{\frac{4}{5}}\begin{pmatrix}
\sin\frac{\pi}{5} & \sin\frac{3\pi}{5}\\
\sin\frac{3\pi}{5} & -\sin\frac{\pi}{5}
\end{pmatrix}.
\end{align}
We will begin with the $G_{2,1}$ CFT, whose partition function is,
\begin{align}
\mathcal{Z}^{G_{2,1}}=\overline{\chi^{G_{2,1}}_0}\chi^{G_{2,1}}_0+\overline{\chi^{G_{2,1}}_{\frac{2}{5}}}\chi^{G_{2,1}}_{\frac{2}{5}}\ ,
\end{align}
where, $\chi^{G_{2,1}}_0,\chi^{G_{2,1}}_{\frac{2}{5}}$ (or $\overline{\chi^{G_{2,1}}_0},\overline{\chi^{G_{2,1}}_{\frac{2}{5}}}$) are (anti-)chiral characters corresponding to
Ka\v{c}-Moody primaries with scaling dimensions $h_L=h_R=0$ and
$h_L=h_R=\frac{2}{5}$ respectively. Their $q$ expansion reads,  
\begin{align}
\chi^{G_{2,1}}_0&=q^{-\frac{7}{60}}(1+14q+42q^2+140q^3+\cdots)\ ,\nonumber\\
\chi^{G_{2,1}}_{\frac{2}{5}}&=q^{\frac{23}{60}}(7+34q+119q^2+322q^3+\cdots)\ .
\end{align}

Let us consider the Verlinde line corresponding to the non-identity
primary. The action of the Verlinde line is given by,
\begin{align} 
\mathcal{L}_{7}|\phi_0\rangle
&=\frac{S_{10}}{S_{00}}|\phi_0\rangle=\alpha|\phi_0\rangle,\nonumber\\
\mathcal{L}_{7}|\phi_{\frac{2}{5}}\rangle
&=\frac{S_{11}}{S_{01}}|\phi_{\frac{2}{5}}\rangle=-\frac{1}{\alpha}
|\phi_{\frac{2}{5}}\rangle,
\end{align}
where $\alpha=\frac{1}{2}(1+\sqrt{5})$ is the golden ratio.  Therefore
the partition function with $\mathcal{L}_7$ insertion is given by,
\begin{align}
  \mathcal{Z}^{G_{2,1}\,\mathcal{L}_7}=\alpha|\chi^{G_{2,1}}_0|^2
  -\frac{1}{\alpha} |\chi^{G_{2,1}}_{\frac{2}{5}}|^2.
\end{align}
The defect Hilbert space partition function is obtained by performing the
modular $S$ transformation.
\begin{align}
\mathcal{Z}^{G_{2,1}}_{\mathcal{L}_7}(\tau)
&=\mathcal{Z}^{G_{2,1}\,\mathcal{L}_7}(-1/\tau)\nonumber\\
&=\alpha|\chi^{G_{2,1}}_0(-1/\tau)|^2-\frac{1}{\alpha}| \chi^{G_{2,1}}_{\frac{2}{5}}
(-1/\tau)|^2\nonumber\\
&=\overline{\chi^{G_{2,1}}_{\frac{2}{5}}}\chi^{G_{2,1}}_{\frac{2}{5}}+\overline{\chi^{G_{2,1}}_0}\chi^{G_{2,1}}_{\frac{2}{5}}
+\overline{\chi^{G_{2,1}}_{\frac{2}{5}}}\chi^{G_{2,1}}_0.
\end{align}
Therefore, in this case, the scaling dimensions $(h_L,h_R)$ of the
operators in the defect Hilbert space is given as
$({\frac{2}{5}},{\frac{2}{5}})$,$(0,\frac{2}{5})$ and
$(\frac{2}{5},0)$. This defect is a non-invertible defect. To see this
note that the fusion rule for the $G_{2,1}$ WZW model dictates the
following fusion rule on $\mathcal{L}_7$,
\begin{align}\label{fusion-rule-G2}
\mathcal{L}_7\times \mathcal{L}_7 = \mathbb{1} + \mathcal{L}_7.
\end{align}
Therefore the defect does not have an inverse under the fusion rule as there is no other primary operator to act as the inverse. 

Just as in the previous sub-sections, the discussion for the defect
$\mathcal{L}_7$ in $G_{2,1}$ carries over to the defect
$\mathcal{L}_{26}$ in $F_{4,1}$ on the replacement of $\chi^{G_{2,1}}$
by the corresponding $\tilde{\chi}^{F_{4,1}}$ whose $q$ expansions
read,
\begin{align}
\tilde{\chi}^{F_{4,1}}_0&=q^{-\frac{13}{60}}(1+52q+377q^2+1976q^3+\cdots)\ ,\nonumber\\
\tilde{\chi}^{F_{4,1}}_{\frac{3}{5}}&=q^{\frac{23}{60}}(26+299q+1702q^2+7475q^3+\cdots)\ .
\end{align}
The character of $E_{8,1}$ is now given as,
\begin{align}
\chi_0^{E_{8,1}}=\chi^{G_{2,1}}_0\tilde{\chi}^{F_{4,1}}_0+\chi^{G_{2,1}}_{\frac{2}{5}}\tilde{\chi}^{F_{4,1}}_{\frac{3}{5}},
\end{align}
In this case we need the following branching rules \cite{mckay1981tables},
\begin{align}
\mathbf{248}&=\mathbf{(14,1)}+\mathbf{(1,52)}+\mathbf{(7,26)}\nonumber\\ \mathbf{3875}&=\mathbf{(7,273)}+\mathbf{(14,52)}+\mathbf{(1,324)}+\mathbf{(27,26)}+\mathbf{(7,26)}\nonumber\\
  &\hspace{5mm}+\mathbf{(27,1)}+\mathbf{(1,1)}.
\end{align}
The defect $\eta_{\frac{14}{5},\frac{26}{5}}$ insertion is given as follows,
\begin{align}
(\chi^{\eta_{\frac{14}{5},\frac{26}{5}}})^{E_{8,1}}&=\alpha\chi^{G_{2,1}}_0\tilde{\chi}^{F_{4,1}}_0-\frac{1}{\alpha}\chi^{G_{2,1}}_\frac{1}{3}\tilde{\chi}^{F_{4,1}}_\frac{2}{3}\nonumber\\
&=q^{-\frac{1}{3}}(1+(124-58\sqrt{5})q+(2062-915\sqrt{5})q^2+\cdots). 
\end{align}
Using the decomposition in terms of the characters above and using the branching rules, we deduce that the representations which receive a contribution of $-\frac{1}{\alpha}$ from the insertion are,
\begin{align}
\mathbf{(7,273)}+\mathbf{(7,26)}+\mathbf{(27,26)}+\mathbf{(7,26)},
\end{align}
while the rest of the representations carry a factor of $\alpha$. However, the Clebsch-Gordan formula for the representation $\mathbf{7}$ in $G_{2}$ reads,
\begin{align}
\mathbf{7}\otimes \mathbf{7}=\mathbf{1}\oplus \mathbf{7} \oplus \mathbf{14} \oplus \mathbf{27}.
\end{align}
Thus, already at the level of the $G_{2,1}$ CFT, one can not explain the insertion of the factors in terms of the Clebsch Gordan formula as we did earlier. In fact, it is this curious feature of $G_2$ that leads to the defect fusion rule \eqref{fusion-rule-G2}. Unlike the cases considered so far where, at level one, only the identity survived in the product of the fundamental and anti-fundamental representation, here the fundamental representation itself appears at the RHS making the defect non-invertible. In this way we get the first instance of a non-invertible defect inside the $E_{8,1}$ WZW CFT.  Although we have indicated the decomposition inside $E_8$ which would give the character with defect insertion, further investigation is needed to obtain a clear understanding of this defect inside $E_{8,1}$ CFT.  One can check that the defect partition function,
\begin{align}
(\mathcal{Z}^{\eta_{\frac{14}{5},\frac{26}{5}}})^{E_{8,1}}=|(\chi^{\eta_{\frac{14}{5},\frac{26}{5}}})^{E_{8,1}}|^2&=|\chi^{G_{2,1}}_{\frac{2}{5}}\tilde{\chi}^{F_{4,1}}_{\frac{3}{5}}+\tilde{\chi}^{F_{4,1}}_0\chi^{G_{2,1}}_{\frac{2}{5}}
+\tilde{\chi}^{F_{4,1}}_{\frac{3}{5}}\chi^{G_{2,1}}_0|^2,
\end{align}
is invariant under $\Gamma_0(5)$.

\subsection{Lee-Yang model CFT($A_{0.5}$)/$E_{7.5}$ IVOA}

For the $A_{0.5}$ CFT and the $E_{7.5}$ IVOA, the central charges are $c_{eff}=\frac{2}{5}$($c=c_{eff}-h=-\frac{22}{5}$) and $c=\frac{38}{5}$. The modular S matrix for $A_{0.5}$ is given by,
\begin{align}
S=\sqrt{\frac{4}{5}}\begin{pmatrix}
-\sin\frac{3\pi}{5} & \sin\frac{\pi}{5}\\
\sin\frac{\pi}{5} & \sin\frac{3\pi}{5}
\end{pmatrix}\ ,
\end{align}
and the partition function is given as,
\begin{align}
\mathcal{Z}=\overline{\chi^{\text{$A_{0.5}$}}}_0\chi^{\text{$A_{0.5}$}}_0+\overline{\chi^{\text{$A_{0.5}$}}}_{-\frac{1}{5}}\chi^\text{$A_{0.5}$}_{-\frac{1}{5}}\ ,
\end{align}
where $\chi^{A_{0.5}}_0,\chi^{A_{0.5}}_{-\frac{1}{5}}$ are characters corresponding to primaries with scaling dimensions $h_L=h_R=0$ and
$h_L=h_R=-\frac{1}{5}$ respectively.  Let us consider the Verlinde
line corresponding to the second primary. The action of the Verlinde
line is given by,
\begin{align}
  \mathcal{L}_{-\frac{1}{5}}|\phi_0\rangle
&=\frac{S_{10}}{S_{00}}|\phi_0\rangle=-\frac{1}{\alpha}|\phi_0\rangle,
\nonumber\\	
\mathcal{L}_{-\frac{1}{5}}|\phi_{-\frac{1}{5}}\rangle
&=\frac{S_{11}}{S_{01}}|\phi_{-\frac{1}{5}}\rangle=\alpha|
\phi_{-\frac{1}{5}}\rangle\ ,
\end{align}
where $\alpha=\frac{1}{2}(1+\sqrt{5})$ is the golden ratio.  Therefore
the partition function with $\mathcal{L}_{-\frac{1}{5}}$ insertion is given by,
\begin{align}
\mathcal{Z}^{A_{0.5}\,\mathcal{L}_{-\frac{1}{5}}}=-\frac{1}{\alpha}|\chi^{\text{$A_{0.5}$}}_0|^2+\alpha|\chi^{\text{$A_{0.5}$}}_{-\frac{1}{5}}|^2.
\end{align}
To obtain the defect Hilbert space partition function, we perform an
$S$ modular transformation.
\begin{align}
\mathcal{Z}^{A_{0.5}}_{\mathcal{L}_{-\frac{1}{5}}}(\tau)
&=\mathcal{Z}^{A_{0.5}\,\mathcal{L}_{-\frac{1}{5}}}(-1/\tau)\nonumber\\
&=-\frac{1}{\alpha}|\chi^{A_{0.5}}_0(-1/\tau)|^2+\alpha
|\chi^{\text{$A_{0.5}$}}_{-\frac{1}{5}}(-1/\tau)|^2\nonumber\\
&=\overline{\chi^{\text{$A_{0.5}$}}_{-\frac{1}{5}}}\chi^{\text{$A_{0.5}$}}_{-\frac{1}{5}}+\overline{\chi^{\text{$A_{0.5}$}}_0}
\chi^{\text{$A_{0.5}$}}_{-\frac{1}{5}}+\overline{\chi^{\text{$A_{0.5}$}}_{-\frac{1}{5}}}\chi^{\text{$A_{0.5}$}}_0.
\end{align}
Therefore the scaling dimensions $(h_L,h_R)$ of the operators in the
defect Hilbert space is given as
$(-{\frac{1}{5}},-{\frac{1}{5}})$,$(0,-\frac{1}{5})$ and
$(-\frac{1}{5},0)$ \cite{Chang:2018iay}. We can similary construct the defect for the $E_{7.5}$ IVOA.

The defect insertion character inside $E_{8,1}$ is given as,
\begin{align}
(\chi^{\eta_{\frac{2}{5},\frac{38}{5}}})^{E_{8,1}}=-\frac{1}{\alpha}\chi^{\text{$A_{0.5}$}}_0\tilde{\chi}^{E_{7.5}}_0+\alpha\chi^{\text{$A_{0.5}$}}_{-\frac{1}{5}}\tilde{\chi}^{E_{7.5}}_\frac{4}{5},
\end{align}
which is a non-invertible defect. The defect Hilbert space partition function,
\begin{align}
(\mathcal{Z}^{\eta_{\frac{2}{5},\frac{38}{5}}})^{E_{8,1}}=	|(\chi_{\eta_{\frac{2}{5},\frac{38}{5}}})^{E_{8,1}}|^2&=|\chi^{\text{$A_{0.5}$}}_{-\frac{1}{5}}\tilde{\chi}^{E_{7.5}}_{\frac{4}{5}}+
\chi^{\text{$A_{0.5}$}}_{-\frac{1}{5}}\tilde{\chi}^{E_{7.5}}_0+\chi^{\text{$A_{0.5}$}}_0\tilde{\chi}^{E_{7.5}}_\frac{4}{5}|^2
\end{align}
is invariant under $\Gamma_0(5)$ \footnote{Note that for the moular
  transformation purpose, one should treat the primary in the Lee-Yang
  model with dimension $-\frac{1}{5}$ as a primary in the $A_{0.5}$
  CFT with effective dimension $\frac{1}{5}$.}.

\subsection{$D_{4,1}$ WZW CFT}
For $D_{4,1}$ theory the central charge is $c=4$. The theory is self-dual inside the $E_{8,1}$ CFT. The modular S matrix is given by,
\begin{align}
S=\frac{1}{2}\begin{pmatrix}
1 & 1 & 1 & 1\\
1 & 1 & -1 & -1\\
1 & -1 & 1 & -1\\
1 & -1 & -1 & 1
\end{pmatrix}.
\end{align}
Partition function is given as,
\begin{align}
\mathcal{Z}^{D_{4,1}}=\overline{\chi^{D_{4,1}}_0}\chi^{D_{4,1}}_0+3\overline{\chi^{D_{4,1}}_{\frac{1}{2}}}\chi^{D_{4,1}}_{\frac{1}{2}}\ ,
\end{align}
where $\chi^{D_{4,1}}_0,\chi^{D_{4,1}}_{\frac{1}{2}}$ are characters corresponding
to Ka\v{c}-Moody primaries with scaling dimensions $h_L=h_R=0$ and
$h_L=h_R=\frac{1}{2}$ respectively.  Let us consider the
Verlinde line corresponding to the second primary. The action
of the Verlinde line is given by,
\begin{align}
\mathcal{L}_v|\phi_0\rangle&=\frac{S_{10}}{S_{00}}|\phi_0\rangle=|\phi_0\rangle,\nonumber\\
\mathcal{L}_v|\phi_v\rangle&=\frac{S_{11}}{S_{01}}|\phi_v\rangle=|\phi_v\rangle,\nonumber\\
\mathcal{L}_v|\phi_s\rangle&=\frac{S_{12}}{S_{02}}|\phi_s\rangle=-|\phi_s\rangle,\nonumber\\
\mathcal{L}_v|\phi_c\rangle&=\frac{S_{13}}{S_{03}}|\phi_c\rangle=-|\phi_c\rangle.
\end{align}
Therefore the partition function with $\mathcal{L}_v$ insertion is given by,
\begin{align}
\mathcal{Z}^{D_{4,1}\,\mathcal{L}_v}=|\chi^{D_{4,1}}_0|^2-|\chi^{D_{4,1}}_{\frac{1}{2}}|^2.
\end{align}
To obtain the defect Hilbert space partition function, we
perform an $S$ modular transformation.
\begin{align}
\mathcal{Z}^{D_{4,1}}_{\mathcal{L}_v}(\tau)
&=\mathcal{Z}^{D_{4,1}\, \mathcal{L}_v}(-1/\tau)\nonumber\\
&=|\chi^{D_{4,1}}_0(-1/\tau)|^2-|\chi^{D_{4,1}}_{\frac{1}{2}}(-1/\tau)|^2\nonumber\\
&=2\overline{\chi^{D_{4,1}}_{\frac{1}{2}}}\chi^{D_{4,1}}_{\frac{1}{2}}+\overline{\chi^{D_{4,1}}_0}
\chi^{D_{4,1}}_{\frac{1}{2}} +\overline{\chi^{D_{4,1}}_{\frac{1}{2}}}\chi^{D_{4,1}}_0.
\end{align}
Therefore the scaling dimensions $(h_L,h_R)$ of the operators in the
defect Hilbert space is given as $({\frac{1}{2}},{\frac{1}{2}})$,
$({\frac{1}{2}},{\frac{1}{2}})$, $(0,\frac{1}{2})$ and
$(\frac{1}{2},0)$. This defect is evidently a $\mathbb{Z}_2\times\mathbb{Z}_2$ defect seen from the action on the primaries above.

The defect insertion character in $E_{8,1}$ is,
\begin{align}
(\chi^{\eta_{4,4}})^{E_{8,1}}=\left(\chi^{D_{4,1}}_0\right)^2-\left(\chi^{D_{4,1}}_{\frac{1}{2}}\right)^2.
\end{align} 
The defect Hilbert space partition function,
\begin{align}
|(\chi^{\eta_{4,4}})^{E_{8,1}}|^2=2\chi^{D_{4,1}}_{\frac{1}{2}}(\chi^{D_{4,1}}_0+\chi^{D_{4,1}}_{\frac{1}{2}}),
\end{align}	
is invariant under $\Gamma_0(2)$. 

\section{Topological defects for $c=24$ Meromorphic CFTs}
In this section, we will briefly discuss how to write down the twisted
characters for $c=24$ meromorphic CFTs which contain commutant pairs
involving one the members from the MMS series CFTs 
\cite{Gaberdiel:2016zke}. In Table-\ref{Table-Meromorphic}, the pairs
are listed where $c,h$ and $\tilde{c},\tilde{h}$ are the central
charge and the conformal dimension of the non-identity primary in the
MMS series and its commutant dual respectively.  The entry
$\tilde{m}_1$ denotes the number of currents in the dual commutant
theory and $M$ denotes the number of currents in the corresponding
$c=24$ meromorphic CFT.
\begin{table}[h]
\begin{center}
	\begin{tabular}{ |c| c| c| c| c |c| c|}
		\hline
		MMS CFT & $c$ & $h$  & $\tilde{c}$& $\tilde{h}$ &  $\tilde{m}_1$ & $M$\\
		\hline
		$A_{1,1}$ & $1$ & $\frac{1}{4}$  & $23$ & $\frac{7}{4}$ & $69$ & $72$\\    
		\hline
		$A_{2,1}$ & $2$ & $\frac{1}{3}$  & $22$ & $\frac{5}{3}$ & $88$ & $96$\\
		\hline
		$G_{2,1}$ & $\frac{14}{5}$ & $\frac{2}{5}$  & $\frac{106}{5}$ & $\frac{8}{5}$ & $106$ & $120$\\
		\hline
		$D_{4,1}$ & $4$ & $\frac{1}{2}$  & $20$ & $\frac{3}{2}$ & $140$ & $168$\\
		\hline
		$F_{4,1}$ & $\frac{26}{5}$ & $\frac{3}{5}$  & $\frac{94}{5}$ & $\frac{7}{5}$ & $188$ & $240$\\
		\hline
		$E_{6,1}$ & $6$ & $\frac{2}{3}$  & $18$ & $\frac{4}{3}$ & $234$ & $312$\\
		\hline
		$E_{7,1}$ & $7$ & $\frac{3}{4}$  & $17$ & $\frac{5}{4}$ & $323$ & $456$\\
		\hline
	\end{tabular}
\end{center}
  \caption{MMS Series of CFTs and their commutant pairs in $c=24$ CFTs.}
  \label{Table-Meromorphic}
\end{table}

As observed in \cite{Gaberdiel:2016zke}, the conformal dimensions of the pairs add upto $2$. The characters above once again satisfy the relation,
\begin{align}
\chi^{\text{Meromorphic}}=j-744+M=\chi_0\tilde{\chi}_0+N\chi_1\tilde{\chi}_1,
\end{align}
where $N$ denotes the degeneracy at the first excited level.

The commutant pair of the MMS series CFT has the following expressions for its characters \cite{Gaberdiel:2016zke},

\begin{align}
	\chi_{0}(\tau)&=j^\frac{\tilde{c}}{24}{ }_{2} F_{1}(-\frac{1}{12}-\frac{\tilde{h}}{2},\frac{7}{12}-\frac{\tilde{h}}{2}, 1-\tilde{h} ;\frac{1728}{j})\ ,\nonumber\\
	\chi_{\tilde{h}}(\tau)&=\frac{|\sqrt{\tilde{m}}|}{\sqrt{N}}j^{\frac{\tilde{c}}{24}-\tilde{h}}{ }_{2} F_{1}(-\frac{1}{12}+\frac{\tilde{h}}{2},\frac{7}{12}+\frac{\tilde{h}}{2}, 1+\tilde{h} ;\frac{1728}{j}),
\end{align}
where,
\begin{align}
	\sqrt{\tilde{m}}=(1728)^{\tilde{h}}\left(\frac{\sin(\pi(\frac{1}{12}+\frac{\tilde{h}}{2}))\sin(\pi(\frac{7}{12}-\frac{\tilde{h}}{2}))}{\sin(\pi(\frac{1}{12}-\frac{\tilde{h}}{2}))\sin(\pi(\frac{7}{12}+\frac{\tilde{h}}{2}))}\right)^{0.5}\frac{\Gamma(1-\tilde{h})\Gamma(\frac{13}{12}+\frac{\tilde{h}}{2})\Gamma(\frac{5}{12}+\frac{\tilde{h}}{2})}{\Gamma(1+\tilde{h})\Gamma(\frac{13}{12}-\frac{\tilde{h}}{2})\Gamma(\frac{5}{12}-\frac{\tilde{h}}{2})}.
\end{align} 
 
In what follows, we will consider defects in a couple of $c=24$ CFTs,
however, our results can be easily generalised to other $c=24$ CFTs
listed in table-\ref{Table-Meromorphic}.  We will discuss the defects
in the $c=24$ CFTs with $M=72$ and $M=456$, and make some comments on
their interpretation. For the $M=72$ CFT, the pair consists of
$A_{1,1}$ CFT whose characters were given in
\eqref{q-expansion-A11}. The corresponding $c=23$ characters have the
$q$ expansion,
\begin{align}
\tilde{\chi}^{c=23}_0&=q^{-\frac{23}{24}}(1+69q+131905q^2+\cdots),\nonumber\\
\tilde{\chi}^{c=23}_\frac{7}{4}&=q^{\frac{19}{24}}(32384+23493120q^2+\cdots).
\end{align}
Using the replacement rule on \eqref{etainsertionsu2}, the defect insertion chracter for the $c=24$ CFT reads,
\begin{align}
(\chi^{\eta_{1,23}})^{c=24}&=\chi^{A_{1,1}}_0\tilde{\chi}^{c=23}_0-\chi^{A_{1,1}}_{\frac{1}{4}}\tilde{\chi}^{c=23}_\frac{7}{4}\nonumber\\
&=\frac{1}{q}+72+67348q+\cdots\ .
\end{align}
Thus we see that the defect does not alter the dimension one
contribution to the character which counts the number of
currents. This is so because the conformal dimensions of the
nontrivial primaries in the commutant pairs add up to 2 and hence do
not contribute at dimension one.  The commutant pairs of CFTs can be
embedded in multiple $c=24$ CFTs, the list of which is given in
\cite{Schellekens:1992db}. For $M=72$ considered above, one of the
possibilities for the current algebra is $(A_{1,1})^{24}$ which is the
15th entry in the list. Here we can see that the above defect can be
interpreted as the center of either $A_{1,1}$ from the MMS series or
the center of the $(A_{1,1})^{23}$ which is the centraliser of the
former inside the $c=24$ CFT.  In the Schellekens classification, this
commutant pair can, in fact, be embedded in any of the entries between
15 and 21 \cite{Schellekens:1992db,Gaberdiel:2016zke}.

Let us now consider the example of $c=24$ CFT with $M=456$.  We considered the characters of the $E_{7,1}$ CFT in \eqref{q-expansion-E71}. The commutant dual has characters,
\begin{align}
\tilde{\chi}_0^{c=17}&=q^{-\frac{17}{24}}(1+323q+60860 q^2+\cdots ),\nonumber\\
\tilde{\chi}_\frac{5}{4}^{c=17}&=q^{\frac{13}{24}}(1632+162656q+4681120 q^2+\cdots).
\end{align}
Therefore the defect insertion character in the $c=24$ CFT is,
\begin{align}
(\chi^{\eta_{7,17}})^{c=24}&=\chi^{E_{7,1}}_0\tilde{\chi}^{c=17}_0-\chi^{E_{7,1}}_{\frac{3}{4}}\tilde{\chi}^{c=17}_\frac{5}{4}\nonumber\\
&=\frac{1}{q}+456+14100q+\cdots\ .
\end{align}
As expected the defect leaves the dimension one contribution which is $456$ unchanged.  In both the cases above, the defect, however, changes the dimension two contributions.  It is straightforward to check that analogous statements hold true for other $c=24$ CFTs listed in the table-\ref{Table-Meromorphic}.

\section{Discussion}
We have studied the topological defect lines in certain 2 character
rational conformal field theories and used them to construct the
defect partition function in $E_{8,1}$ conformal field theory.  We
used the MMS series of CFTs, which form commutant pairs in the
$E_{8,1}$ conformal field theory for this purpose.  Using the TDLs of
the MMS series, we construct the defect characters of the $E_{8,1}$
theory for which we give a group theoretic interpretation.  These
novel defects do not preserve full $E_8$ symmetry, which shows up as
the reduction in the degeneracies at various levels in the identity
character.  In particular, the reduction in the degeneracy at level
one of the identity character corresponds to the symmetries preserved
by the defect.  This symmetry is consistent with our group theoretic
understanding of the defect partition functions.

We also analysed the commutant pairs of the MMS CFTs inside $c=24$
meromorphic CFTs.  We found that the defects preserve dimension 1
symmetries of $c=24$ meromorphic CFTs.  We attributed this
contrasting behaviour of the TDLs to their embedding in dimension 1
operators in the case of $E_8$ theory, and in dimension 2 operators in case
of $c=24$ CFTs.  It can be confirmed by seeing that the defect does
reduce the degeneracy at level 2 in $c=24$ CFTs.  Although we have
not done an exhaustive study of defects in the commutant pairs in $c=24$
theories, our results can be generalised to other pairs in a
straightforward manner.

It would be interesting to generalise this method of deriving defect
characters and partition functions by using three character conformal
field theories.  It would be curious to see how this formalism generalises to
triple commutant or multiple commutant cases.  Recently novel coset relations
have been proposed between the four-point functions of the currents in
the $E_{8,1}$ theory and conformal blocks in the commutant pairs
inside $E_{8,1}$ \cite{Mukhi:2020sxt}.  It will be interesting to see
the consequence of the defect partition functions derived here, on
these four-point function of currents.  We hope to address some of
these questions in the future.

The Lie groups that appear in the MMS series are also referred to as
the Deligne-Cvitanovic series of exceptional Lie
groups\cite{cvitanovic1977classical,deligne1996serie,deligne1996serie2,cvitanovic2008group}.
These exceptional Lie groups appear in the study of the Higgs branch
of four dimensional $\mathcal{N}=2$ super-Yang-Mills theories\cite{beem:2013sza}.  The
corresponding two dimensional CFTs have current algebra symmetries
belonging to the Deligne-Cvitanovic series of exceptional Lie groups.
Although these two dimensional theories have negative central charge
and hence are nonunitary theories, they are also two character
theories and possess similar commutant pair relationships.  Our
results when applied to these theories shed light on the spectrum of
line defects in the four dimensional rank 1 super-Yang-Mills
theories\cite{abhishek2021}.

\acknowledgments We thank Pramath A V for many useful discussions. We
thank Ratul Mahanta for collaboration in the early stages of the
project and discussions.  DPJ acknowledges support from SERB grant
CRG/2018/002835.

\bibliography{mms-defects.bib}

\end{document}